\documentclass[twocolumn,prl,aps]{revtex4}
\usepackage{epsfig}
\usepackage{amsmath, amsthm, amsfonts}

\newcommand\beq{\begin{equation}}
\newcommand\eeq{\end{equation}}
\newcommand\bea{\begin{eqnarray}}
\newcommand\eea{\end{eqnarray}}

\newcommand{\ket}[1]{| #1 \rangle}
\newcommand{\bra}[1]{\langle #1 |}

\newcommand{\ba}{\begin{array}}
\newcommand{\ea}{\end{array}}

\begin{document}

\title{Local Hidden Variable Theories for Quantum States}

\author{Barbara M. Terhal$^{1,2}$, Andrew C. Doherty$^2$ and David
  Schwab$^3$}
\affiliation{$^1$ IBM Watson Research Center, P.O.
  Box 218, Yorktown Heights, NY 10598, USA, \\
  $^2$ Institute for Quantum Information, Caltech 107--81,
  Pasadena, CA 91125, USA,\\
  $^3$ Department of Physics, Cornell University, 109 Clark Hall,
  Ithaca, NY 14853, USA} \date{\today}

\begin{abstract}
  While all bipartite pure entangled states violate some Bell
  inequality, the relationship between entanglement and non-locality
  for mixed quantum states is not well understood. We introduce a
  simple and efficient algorithmic approach for the problem of
  constructing local hidden variable theories for quantum states. The
  method is based on constructing a so-called symmetric
  quasi-extension of the quantum state that gives rise to a local
  hidden variable model with a certain numbers of settings for the
  observers Alice and Bob.
  We use this method to analytically construct local hidden variable
  theories for any bound entangled state based on a real unextendible
  product basis (UPB) with two measurement settings for Alice and Bob.
  The problem can be approached by semi-definite programming and we
  present our numerical and analytical results for various classes of
  states.
\end{abstract}



\maketitle


It was John Bell \cite{bell:epr} who quantified how measurements on
entangled quantum mechanical systems can invalidate local classical
models of reality. His original inequality has generated a field of
research devoted to general Bell inequalities
and experimentally observed violations of such inequalities. \\
\indent Perhaps surprisingly, the nature of the set of states that
violate local realism is poorly understood, although it is known
from the seminal work of Werner~\cite{werner:lhv} that not all
entangled states violate a Bell inequality. Recent results in
quantum information theory have revealed the complex structure of
the set of entangled states but have as yet shed little light on
the relation between this structure and violation of Bell
inequalities. For example, it has been conjectured by Peres
\cite{peres:bell} that so called bound entangled states which
satisfy the Peres-Horodecki ``partial transposition'' criterion
\cite{HHH:PPT}, do not violate any Bell inequalities. There are
various results that support this conjecture both in the bipartite
and multipartite case, see Refs.
\cite{WW:bell,WW:bellmulti,dur:bell,ASW:bell,acin+:qnl,KZG:lhv},
but none of the results is conclusive.
What has been lacking in the literature so far is a systematic way of
deciding whether a quantum state does or does not violate some Bell
inequality. The difficulty is that the possible types of local
measurements and the number of measurements that observers can perform
is in principle unbounded and the enumeration of Bell
inequalities is computationally hard \cite{pitowsky:polytope}. \\
\indent In this Letter we present the first systematic approach for
constructing local hidden variable theories for quantum states,
depending only on the number of local measurement settings for each
observer. Our approach has yielded both numerically constructed local
hidden variable theories for a variety of quantum states as well as
analytical results for Werner states \cite{werner:lhv} and a class of
bound entangled states based on
real UPBs \cite{bennett+:upb}. \\
\indent Before we can state our main result, we recapitulate the
mathematics of local hidden variable (LHV) models and Bell
inequalities for bipartite systems
\footnote{In the literature one encounters both a weak as well as a
  strong version of a local hidden variable test. In the weak version
  the goal is to reproduce the {\em expectation value} of a set of
  local observables on a particular quantum system by means of a local
  hidden variable model. In the strong version one attempts to
  reproduce the exact probabilities of outcome for a set of local
  measurements by means of a local hidden variable model.  All the
  results in this paper hold for the stronger model.}. We refer the
reader to Ref. \cite{peres:bell, pitowsky:polytope, WW:overview} for
some literature on the theoretical formulation of general Bell
inequalities. Each of the observers, Alice and Bob, has a set of local
measurements. Let $i=1, \ldots, s_a$ be the number of measurements for
Alice and let each measurement have $o_a(i)$ outcomes. Let $k=1,
\ldots ,s_b$ be the number of measurements for Bob and $o_b(k)$ be the
number of outcomes per measurement. The probability $P_{ij,kl}$
denotes the probability that Alice's $i$th measurement has outcome $j$
and Bob's $k$th measurement has outcome $l$. A local hidden variable
model assumes the existence of a shared random variable between Alice
and Bob that is used to locally generate a measurement outcome
depending only on the choice of the local measurement (and not on the
choice of the other, remote, measurement). The local hidden variable
model generates the probability vector $\vec{P}$ with entries
$P_{ij,kl}$ when it generates measurement outcomes in accordance with
these probabilities. \indent Mathematically one defines a convex set
$S(s_a,s_b,o_a,o_b)$ which is the set of probability vectors $\vec{P}$
that can be generated by LHV models. It is known that $S$ is a
polytope and that the extremal vectors $\vec{B}$ of $S$ are vectors
with $0,1$ entries~\cite{WW:overview}.
For more information on polytopes, see for
example~\cite{book:ziegler}. These extremal vectors $\vec{B}$
correspond to the situation in which the outcomes of the measurements
are determined with certainty and can be labelled by 2 sets of indices
${\bf m}=(m_1, \ldots,m_{s_a})$ where $m_i=1, \ldots,o_a(i)$ and ${\bf
  n}=(n_1, \ldots,n_{s_b})$ where $n_k=1, \ldots,o_b(k)$. A brief
expression for these extremal vectors is \beq B_{ij,kl}^{{\bf m},{\bf
    n}}=\delta_{j{m}_i}\delta_{l{n}_k}.
\label{defb} \eeq In words, each extremal vector specifies a
single outcome with probability one for each local measurement,
independently of the measurement made by the other parties. An example
of an extremal vector for $s_a=s_b=2$ and $o_a=o_b=3$ is given in
Table \ref{table1}.

For a quantum mechanical system $\rho$ in ${\cal H}_{d_A} \otimes
{\cal H}_{d_B}$ the probability $P_{ij,kl}$ is given by
$P_{ij,kl}(\rho)={\rm Tr}\, E_{ij}^{A} \otimes E_{kl}^B \rho$.  Here
$\{E_{ij}^A \geq 0:\sum_j E_{ij}^A=I_{d_A}\}$ are the POVM elements
for Alice's $i$th measurement and $\{E_{kl}^B\}$ are the POVM elements
for Bob's $k$th measurement. There is a violation of a Bell inequality
if and only if $P_{ij,kl}$ cannot be generated by a LHV model, or
$\vec{P} \not \in S$.

\begin{table}[h]
\begin{tabular}{r|c c c|c c c}
        &   & $B_1\;\;\;\;$ &  &  & $B_2\;\;\;\;$ &\\
        & 1 & 2 & 3 & 1 & 2 & 3 \\ \hline
       1 & 1 & 0 & 0 & 0 & 1 & 0 \\
      $A_1\;$ 2 & 0 & 0 & 0 & 0 & 0 & 0 \\
       3 & 0 & 0 & 0 & 0 & 0 & 0 \\ \hline
  1 & 0 & 0 & 0 & 0 & 0 & 0 \\
  $A_2\;$      2& 0 & 0 & 0 & 0 & 0 & 0 \\
        3& 1 & 0 & 0 & 0 & 1 & 0 \\
\end{tabular}
\caption{An example of an extremal B-vector for 2 settings for
Alice and Bob with 3 outcomes per setting.}
 \label{table1}
\end{table}

In this Letter we will prove the first necessary condition for a state
to violate a Bell inequality depending only on the number of settings
for Alice and Bob. We will explicitly construct a LHV model in a
$(s_a=2,s_b=2)$ setting for any bound entangled state based on a {\em
  real} unextendible product basis \cite{bennett+:upb}. Then we will
discuss numerical work that shows that many of the known bipartite
bound entangled states cannot violate a Bell inequality with two
settings for Alice and Bob. Finally, we will partially reproduce and
extend some of Werner's original results by showing that it is
possible to use our procedure to analytically construct LHV theories
for Werner states.  It is noteworthy to mention that our methods
(Theorem 1 and Theorem 2) straightforwardly generalize to multipartite
states, even though we have not explored this direction.

We will connect violations of Bell inequalities to the existence of a
symmetric (quasi-) extension of a quantum state \footnote{A connection
  between Bell inequalities and extensions has been made previously by
  R.F. Werner \cite{werner:ext} who used violations of a CHSH
  inequality to show that there exist quantum
  states on AB and A'B with a common reduction on B for which there is
  no joint quantum state on AA'B. In the case where the states on AB
  and A'B are required to be the same, the existence of such a joint
  state implies the existence of a symmetric extension.}. An extension
of a quantum state
$\rho$ on, say, a system $AB$, is another quantum state defined on a
system $ABC$ such that when we trace over $C$ we obtain the original
quantum state $\rho$. We are interested in the situation where the
system $C=A^{\otimes (s_a-1)}\otimes B^{\otimes (s_b-1)}$ and we will
demand that the extension be invariant under all permutations of the
$s_a$ copies of system $A$ among each other and similarly invariant
under any permutation of the $B$ systems. It is clear that if the
quantum state $\rho$ is separable, i.e.  $\rho=\sum_i p_i
(\ket{\psi_i}\bra{\psi_i})_A \otimes (\ket{\phi_i}\bra{\phi_i})_B$,
such an extension always exists: we just copy the individual product
states onto the other spaces: \beq \rho_{ext}=\sum_i p_i
(\ket{\psi_i}\bra{\psi_i})^{\otimes s_a} \otimes
(\ket{\phi_i}\bra{\phi_i})^{\otimes s_b}.  \eeq If the state $\rho$ is
a pure entangled state, then it is also clear that such a symmetric
extension cannot exist. The symmetry requirement implies that the pure
entangled state $\rho_{AB}$ must equal $\rho_{A'B}$ where $A'$ is
another $A$-system, which is impossible. In popular terms we may say
that pure entanglement is `monogamous', $B$ cannot be entangled with
$A$ and $A'$ at the same time. In some sense what we show in this
paper is that (1) a violation of a Bell inequality indicates that the
entanglement in the quantum state is `monogamous' and (2) there are
many mixed entangled states whose entanglement is not monogamous.

Thus the existence of a symmetric extension can be viewed as a
separability criterion (see Ref. \cite{DPS:ext} for a similar but
stronger separability criterion where one demands that the symmetric
extension has positive partial transposes). For considering Bell
inequality violations we generalize our criterion slightly and ask
whether a state has a symmetric {\em quasi-extension} $H_{\rho}$ which
is not necessarily positive. In order to define this notion we need
the definition of a multi-partite entanglement witness, which is an
entanglement witness which can detect any multi-partite entanglement
in a state. It has the property that for all states
$\psi_1,\ldots,\psi_{s_a},\phi_1,\ldots,\phi_{s_b}$,
$\bra{\psi_1,\ldots,\psi_{s_a},\phi_1,\ldots,\phi_{s_b}} H_{\rho}
\ket{\psi_1,\ldots, \psi_{s_a},\phi_1, \ldots,\phi_{s_b}} \geq 0$.

{\em Definition [Symmetric Quasi-Extension]:} Let $\pi:{\mathcal
  H}^{\otimes s}\rightarrow {\mathcal H}^{\otimes s}$ be a permutation
of spaces ${\mathcal H}$ in ${\mathcal H}^{\otimes s}$. We define \beq
{\sf Sym}(\rho)=\frac{1}{s!}  \sum_{\pi} \pi \rho {\pi}^{\dagger}.
\eeq We say that $\rho$ on ${\mathcal H}_A \otimes {\mathcal H}_B$ has
a $(s_a,s_b)$-symmetric quasi-extension when there exists a
multi-partite entanglement witness $H_{\rho}$ on ${\mathcal
  H}_A^{\otimes s_a} \otimes {\mathcal H}_B^{\otimes s_b}$ such that
${\rm Tr}_{{\mathcal H}_A^{\otimes (s_a-1)},{\mathcal H}_B^{\otimes
    (s_b-1)}}H_{\rho}=\rho$ and $H_{\rho}= {\sf Sym}_A
\otimes {\sf Sym}_B(H_{\rho}).$\\
\indent The reason for considering such quasi-extensions is clear
from the following theorems which are the main results of this
Letter.

{\em Theorem 1:} If $\rho$ has a $(s_a,s_b)$-symmetric quasi-extension
then $\rho$
does not violate a Bell inequality with $(s_a,s_b)$ settings. \\

Before proving this theorem, it is important to note the generality of
the result; it holds for all possible choices of measurements which
includes POVM measurements with an unbounded number of measurement
outcomes. We will show below that the quasi-extension of $\rho$
effectively creates a LHV model for $\rho$ when Alice and Bob have
$s_a$ and $s_b$ arbitrary measurements.

{\em Proof} We prove our theorem by extracting an LHV model from the
quasi-extension. The LHV model for $\rho$ for $(s_a,s_b)$ settings
should reproduce the vector $P_{ij,kl}(\rho)={\rm Tr}\,E_{ij}^A
\otimes E_{kl}^B \rho$ for all possible choices of POVM measurements
$\{E_{ij}^A,E_{kl}^B\}$, as a convex combination of the extremal
B-vectors, i.e.
\beq P_{ij,kl}(\rho)=\sum_{{\bf m},{\bf n}} p_{{\bf m},{\bf
    n}}(\{E_{ij}^A,E_{kl}^B\}, \rho) B_{ij,kl}^{{\bf m},{\bf n}},
\label{lhvdef}
\eeq where $p_{{\bf m},{\bf n}}(.) \geq 0$. If a symmetric
quasi-extension exists for $\rho$ then ${\rm Tr}\, E_{ij}^A \otimes
E_{kl}^B \;\rho={\rm Tr}\, (E_{ij}^A \otimes E_{kl}^B\otimes {\bf I})
\;H_{\rho}$. Using the definition of the B-vectors, the properties of
the POVMs
($\sum_j E_{ij}^{A,B}=I_{d_{A,B}}$),
and the symmetry
properties of $H_{\rho}$ it is not hard to verify that \beq
P_{ij,kl}(\rho)={\rm Tr}\, E_{ij}^A \otimes E_{kl}^B
\;\rho=\sum_{\bf{m},\bf{n}} ({\rm Tr}\, {\bf E}_{\bf{m}}^A \otimes
{\bf E}_{\bf{n}}^B H_{\rho}) B_{ij,kl}^{\bf{m},\bf{n}}. \label{qe}
\eeq Here ${\bf E}_{\bf m}^A=E_{1m_1}^A \otimes E_{2m_2}^A
\otimes \ldots \otimes E_{s_am_{s_a}}^A$ and similarly for ${\bf
E}_{\bf n}^B$. Since $H_{\rho}$ is a quasi-extension
$p_{\bf{m},\bf{n}}(\{E_{ij}^A,E_{kl}^B\},\rho) \equiv {\rm Tr} {\bf
  E}_{\bf{m}}^A \otimes {\bf E}_{\bf{n}}^B H_{\rho} \geq 0$, and we
have obtained a LHV model. $\Box$

One way of looking at this result is the following
\cite{priv:wolf}. If $\rho$ has a symmetric extension
$\tilde{\rho}$, then instead of measurement on $\rho$, Alice and
Bob can do measurements on ${\tilde \rho}$. Due to the symmetry
Alice can do the first measurement on the first Alice space and
the second measurement on the second Alice space etc. But now
these measurements are all commuting, and can be considered as one
big measurement. But we know that when Alice and Bob each have
only a single measurement a LHV model for their measurements
exists and thus we have a LHV model for the measurements on
$\rho$.
With this picture in mind, it is not hard to understand the
following strenghtening of our results (see also Ref.
\cite{werner:ext}):

{\em Theorem 2:} If $\rho$ has a $(1,s_b)$-symmetric
quasi-extension then $\rho$ does not
violate a Bell inequality with $s_b$ settings for Bob and any number
of settings for Alice. \\

{\em Remark} The theorem also holds when Alice and Bob are
interchanged.

{\em Proof}: The intuition behind this theorem relies on the fact that
there are no violations of Bell inequalities when one party has only
one measurement setting, thus suggesting that it is unnecessary to
extend to copies of Alice's space as well as Bob's. Here is the local hidden
variable model that we construct from a quasi-extension $H_{\rho}$, on
${\cal H}_A \otimes {\cal H}_B^{\otimes s_b}$. We set
\beq
p_{\bf{m},\bf{n}}((\{E_{ij}^A,E_{kl}^B\},
  \rho))=\frac{\Pi_{i'=1}^{s_a}({\rm Tr} E_{i'm_{i'}}^A
   \otimes
  {\bf E}_{\bf{n}}^B H_{\rho})}{({\rm Tr}\,I_A \otimes E_{\bf
    n}H_{\rho})^{s_a-1}}.
\eeq
Each $p_{\bf{m},\bf{n}}$ is nonnegative since $H_{\rho}$ is an
entanglement witness. We can substitute this expression in Eq.
(\ref{lhvdef}) and verify that we obtain the correct probabilities
$P_{ij,kl}(\rho)$ by using the definition of the B-vectors, the
normalization of the POVMs, and the symmetry of $H_{\rho}$ as
before.  $\Box$


This method for constructing LHV theories may be implemented both
numerically and analytically. Let us first show a simple analytic
construction of a $(2,2)$-symmetric extension for {\em any} bound
entangled state based on a {\em real} unextendible product basis
\cite{bennett+:upb}. Let $P_{BE}=I-\sum_i \ket{a_i,b_i}\bra{a_i,b_i}$
be the projector onto such a bound entangled state, where
$\{\ket{a_i,b_i}=\ket{a_i^*,b_i^*}\}$ is the real unextendible product
basis. Our (unnormalized) extension will be
$\ket{\Psi}_{A_2A_1}\otimes \ket{\Psi}_{B_1B_2}-\sum_i
\ket{a_i,a_i,b_i,b_i}_{A_2A_1B_1B_2}$ where $\ket{\Psi}=\sum_i
\ket{ii}$. It is evident that this extension has the desired symmetry
property. It is not hard to verify that by tracing over the systems
$A_2$ and $B_2$ we obtain $P_{BE}^2=P_{BE}$. The existence of a
symmetric $(2,2)$-extension implies the existence of both $(2,1)$ and
$(1,2)$ symmetric extensions for the state by tracing out copies of
$A$ or $B$, so any Bell
inequality violation for this class of states must involve more than
two measurement settings for both parties.

We have implemented numerical tests for the conditions of these two
theorems. Firstly we look for the existence of a symmetric extension with
$H_{\rho}\geq 0$. If such an extension does not exist, there is still
the possibility that some other kind of quasi-extension does exist. We
have focussed on the existence of a decomposable entanglement witness
$H_{\rho}$ because in both these cases the numerical problem
corresponds to a semi-definite program \cite{VB:sp}. We label the
partitions of $\mathcal{H}_A^{\otimes s_a} \otimes
\mathcal{H}_B^{\otimes s_b}$ into bipartite systems by $p$ and we
denote partial transposition with respect to one of the two subsystems
as $T_p$. A decomposable entanglement witness may then be written as
$H_{\rho}=P+\sum_p Q_p^{T_p}$ where $P \geq 0$, $Q_p \geq 0$ for all
$p$. (In fact, due to the symmetry it is only necessary to include
partitions unrelated by permutations of copies of $A$ or copies of $B$
in the sum, as in Ref.  \cite{DPS:ext}.)

Semi-definite programs correspond to optimizations of linear functions
on positive matrices subject to trace constraints. They are convex
optimizations and are particularly tractable both analytically and
numerically. We show how to numerically construct symmetric
extensions, the decomposable quasi-extension case is very similar. The
condition that the partial trace of $H_{\rho}$ is $\rho$ is equivalent
to requiring that ${\rm Tr}\,(X\otimes {\bf I}) H_{\rho} = {\rm Tr}\,
X \rho$ for all operators $X$ on $\mathcal{H}_A \otimes
\mathcal{H}_B$. If we write $X$ in terms of a basis $\{\sigma_i\}$ for
the vector space of operators then by linearity it is enough to check
that this trace constraint holds for each element of the basis. We
will assume that the basis is orthogonal in the trace inner product
${\rm Tr}\,\sigma_i\sigma_j=\delta_{ij}$ and that
$\sigma_0=I_{d_A}\otimes I_{d_B}/\sqrt{d_A d_B}$. The index $i$ ranges
from zero to $(d_A d_B)^2-1$.
Consider then this semi-definite program
\begin{eqnarray*}
  \label{eq:optim}
  {\rm minimize}\quad &  {\rm Tr}\,K, \\
  {\rm subject\ to}\quad & {\rm Tr}\, {\sf Sym}_A \otimes
  {\sf Sym}_B(\sigma_i \otimes {\bf I})K = r_i, \quad i>0, \\
& K\geq 0,
\end{eqnarray*}
where $r_i={\rm Tr}\,\sigma_i \rho$. If the optimum is less than or
equal to one, then, by adding a multiple of the identity to the
optimal $K$, we obtain some $K_{\rho}$ that satisfies ${\rm
  Tr}\,K_{\rho}=1$ as well as the other constraints. If we define
$H_{\rho}\equiv {\sf Sym}_A \otimes {\sf Sym}_B (K_{\rho})$ it is
clear that $H_{\rho}$ is a $(s_a,s_b)$-symmetric extension of $\rho$.
Duality properties of semi-definite programs imply that an optimum
greater than one precludes the existence of a $(s_a,s_b)$-symmetric
extension \cite{VB:sp}
(see also the Appendix).

We have implemented this semi-definite program using SeDuMi
\cite{SeDuMi} for several examples of bound entangled states with
$d_A=d_B=3$. The results are summarized in Table \ref{numres}.
For $(1,2)$ settings the extension code took 1--3s to run on a 500
MHz Pentium 3 desktop with 500 MB of RAM, while the quasi-extension
code took 1.5--4s. The computation for both forms of the code as
described above should scale roughly as $d_A^{2.5s_a+4}d_B^{2.5s_b+4}$
\cite{VB:sp}. However as $s_a$ and $s_b$ grow, the permutation
symmetry of the extensions can be used to dramatically reduce the size
of the problem by block diagonalizing the matrices ${\sf Sym}_A
\otimes {\sf Sym}_B(\sigma_i \otimes {\bf I})$ and removing repeated
blocks \cite{GP:sss,Rains:spde}. For a fixed $d_A,d_B$ the computation
will scale polynomially with $s_a$ and $s_b$. We implemented such a
code in the case of $(1,3)$-extensions which took 1.5--4s for a
single state.

The Choi-Horodecki (C-H) states considered in Ref.
\cite{HHH:beactivate} depend on a parameter $\alpha$ and include
separable ($\alpha \in [2,3]$), bound entangled ($\alpha \in (3,4]$)
and nonpositive partial transpose states for $\alpha > 4$.  They turn
out to have $(2,1)$-symmetric extensions well into the range for which
the states are entangled. Over the range $\alpha \in [4.34, 4.84]$
they have decomposable symmetric quasi-extensions but no symmetric
extensions showing that the former property provides a strictly
stronger sufficient condition for the existence of an LHV
theory.

On the other hand, we found that the two parameter family of bound
entangled states introduced by Horodecki and Lewenstein
\cite{HL:be} do not have $(2,1)$-symmetric extensions or
quasi-extensions. Also, many of the states described by Bru\ss\
and Peres \cite{BP:be} do not appear to possess symmetric
quasi-extensions. However, for several examples of these two kinds
of states we searched numerically over measurement settings to
look for violations of extremal Bell inequalities for $s_a=s_b=2$
and three outcome measurements, and also $s_a=s_b=3$ and two
outcome measurements, without success. Note that states may have
$(s,1)$-extensions and no $(1,s)$-extensions, we have performed
both tests in all cases. Although this possibility does not affect
the overall conclusion, the states of \cite{BP:be} for example are
sufficiently asymmetric with respect to swapping A and B that for
$s=2$ we found examples having one kind of extension but not the
other, as well as states having both kinds (these are examples
with $(2,2)$-extensions which implies this latter condition).

We found that, although they have $(2,1)$-extensions, only a few
of the general complex UPB states of \cite{divincenzo+:upb} have
$(3,1)$-extensions and similarly the C-H states have
$(3,1)$-extensions for a reduced range of values of $\alpha$. We
did not find examples of Bru\ss-Peres states having
$(3,1)$-extensions.

\begin{table}[h]
\begin{tabular}{r|c c |c|}
        &  $(2,1),(1,2)$   &  & $(3,1),(1,3)$  \\\

        & ext & q-ext  & ext   \\ \hline
      C-H \cite{HHH:beactivate}:$\alpha \in$ & $[2,4.33]$ &
      $[2,4.84]$ & $[2,4.00]$  \\
      Complex UPB  \cite{divincenzo+:upb}  & yes & yes & few\\
      H-L \cite{HL:be} & no & no & no \\
      Bru\ss-Peres \cite{BP:be} & few & few & no \\
Werner  \cite{werner:lhv}       & $d\geq 3$& $d\geq 3$ & $d \geq 4$ \\
      Werner $d=2,\Phi\geq $      & $-1/2$ & $-1/2$ &
      $-1/3$  \\
  \end{tabular}
\caption{Numerical results on the existence of symmetric
extensions (ext) and decomposable quasi-extensions (q-ext) for
$(s_a=1,s_b=2)$, $(s_a=2,s_b=1)$, $(s_a=1,s_b=3)$, and $(s_a=3,s_b=1)$.}
  \label{numres}
\end{table}

Finally we considered Werner states \cite{werner:lhv} defined in
dimensions $d=d_A=d_B\geq 2$ as
$\rho_W=\frac{1}{d^3-d}(I(d-\Phi)+(d\Phi-1)V)$ where $V$ is the
flip operator. Werner \cite{werner:lhv} showed that for $\Phi \geq
-1+ \frac{d+1}{d^2}$ these states do not violate any Bell
inequality with an {\em arbitrary} number $s_a,s_b$ of von Neumann
measurements (in Ref. \cite{barrett:POVM} the author constructs
LHV models for arbitrary {\em POVM} settings for a more restricted
range of $\Phi$). We found that using symmetry techniques similar
to those in Ref. \cite{EW:uuu} it is possible to {\em
analytically} solve the dual optimization problem to the
semi-definite program described above,
see the Appendix. The value of the optimum establishes that {\em
all} Werner states have symmetric extensions so long as
$s_a+s_b\leq d$. Hence these states have LHV theories for all Bell
experiments where the minimum number of settings $s=\min(s_a,s_b)$
satisfies $s+1 \leq d$. This result is more general than Werner's
in the sense that, like in Ref. \cite{barrett:POVM}, it holds for
general POVM elements. It is weaker in the sense that the number
of settings is bounded by the dimension of the space. Numerical
and analytical results (see Table II and the Appendix) show that
Werner states for $d=2$ actually have symmetric (quasi-)extensions
beyond this analytically derived bound.

Even though our method is the most powerful tool to date for
constructing local hidden variable theories, we believe that it is
unlikely that every LHV model can be constructed from a symmetric
quasi-extension. Our work is only the starting point for a more
thorough exploration of the existence of (quasi-)extensions for
entangled quantum states. In particular, it is an intriguing and
open question whether there exist entangled states that have
$(s_a=1, s_b \rightarrow \infty)$ quasi-extensions.  In fact we
heard that it has been proved that only separable states have
$(s_a=1,s_b \rightarrow \infty)$ extensions \cite{priv:SW,
FLV:sym, RW:meanfield}.

We would like to thank Dave Bacon and Ben Toner for interesting
discussions concerning Bell inequalities and for providing us with
extremal Bell inequalities used in some of the numerical work. We
are very grateful to M.M. Wolf for his insightful comments on the
original draft of this paper and for bringing Ref.
\cite{werner:ext} to our attention. BMT and ACD acknowledge
support from the National Science Foundation under Grant. No.
EIA-0086038. ACD acknowledges support from the Caltech MURI Center
for Quantum Networks administered by the Army Research Office
under Grant DAAD19-00-1-0374 and BMT from the National Security
Agency and the Advanced Research and Development Activity through
Army Research Office contract number DAAD19-01-C-0056. ACD thanks
Mark Kasevich and Steve Girvin for their hospitality at Yale
University where part of this work was completed. DS thanks the
IQI for the opportunity to work as a summer student.

\section*{Appendix}
In this Appendix we discuss the semidefinite program that constructs
symmetric extensions in more detail and show that it is possible to
construct symmetric
extensions for many Werner states using our semi-definite programming
approach. The ingredients of the argument are the semi-definite
programming duality, the behavior of convex optimizations under
symmetry and some simple group representation theory for the
permutation group.

The semi-definite program that attempts to construct a symmetric
extension has a natural dual that will turn out to be simpler to
solve. We first review, for completeness, some standard descriptions
 and properties of semi-definite programs and their duals.

Vandenberghe and Boyd \cite{VB:sp} write the general semi-definite
program as
\begin{eqnarray*}
\label{sdp}
{\rm minimize} \quad &  {\bf c}^T {\bf x},  \\
{\rm subject\ to} \quad & F_0 +\sum_i x_i F_i  \geq 0,
\end{eqnarray*}
where ${\bf c}$ is a vector of length $m$, and $(F_0,F_i)$ are
$n$-by-$n$ Hermitian matrices with $i=1,...,m$. The optimization
variables form the vector ${\bf x}$, also of length $m$. If there
exists any vector ${\bf x}$ such that $F({\bf x})=F_0 +\sum_i x_i F_i
> 0$ the semi-definite program is said to be strictly feasible. The
dual optimization, also a semi-definite program, may be written
\begin{eqnarray*}
{\rm maximize} \quad & -{\rm Tr}\,F_0 Z,  \\
{\rm subject\ to} \quad & Z \geq 0,  \\
 & {\rm Tr}\,F_i Z = c_i,
\end{eqnarray*}
where the optimization variable is the $n$-by-$n$ matrix $Z$.  Again,
the dual semi-definite program is said to be strictly feasible if
there is an $Z>0$ satisfying the trace constraints.

The most important relation between the primal and dual optimizations
is that {\em all allowed values of the primal objective function are
  greater than all allowed values of the dual objective function}.
Thus feasible points of the dual problem can be used to bound the
optimum of any semi-definite program. This property results from
simply evaluating the difference between the primal and dual objective
functions for any feasible pair $({\bf x},Z)$
\begin{equation*}
  \label{eq:duality}
  {\bf c}^T {\bf x}+{\rm Tr}\,F_0 Z=\sum_i {\rm Tr}\,x_i F_i Z + {\rm
  Tr}\,F_0 Z= {\rm Tr}\,F({\bf x}) Z \geq 0.
\end{equation*}
The first equality results from the dual feasibility constraints and
the inequality holds since ${\rm Tr}\,AB$ is positive if $A$ and $B$
are positive and Hermitian. Typically, the dual optimizations are more
closely related, this is captured in Theorem 3.1 of \cite{VB:sp}. {\em
  If both the primal and dual forms of a semi-definite program are
  strictly feasible, their optima are equal and achieved by some
  feasible pair $({\bf x}_{\rm opt},Z_{\rm opt})$}.

Comparing the semi-definite programs above with the semi-definite
program for constructing symmetric extensions we see that it
corresponds to the dual form defined above with $F_i={\sf Sym}_A
\otimes {\sf Sym}_B(\sigma_i \otimes {\bf I})$, $c_i=r_i={\rm
  Tr}\,\sigma _i \rho$ and $F_0=I_{d_A} \otimes I_{d_B} \otimes {\bf
  I}$. We have $m=(d_A d_B)^2-1$ and $n=d_A^{s_a}d_B^{s_b}$. The sign
of the objective function was changed for clarity. Recall that there
is a symmetric extension for $\rho$ so long as the optimum is less
than or equal to one. From now on we will write ${\sf Sym'}={\sf
  Sym}_A \otimes {\sf Sym}_B$. To see that this optimization is
strictly feasible consider any, not necessarily positive, matrix $K'$
satisfying the trace constraints ($K'$ exists since these constraints
fix only a small number of the matrix elements of $K'$). Then
$K=K'+\eta I_{d_A}\otimes I_{d_B} \otimes {\bf I}$ is strictly
positive so long as $\eta>0$ is greater than the magnitude of the
largest negative eigenvalue of $K'$. $K$ defined in this way still
satisfies the trace constraints so the semi-definite program is
strictly feasible. If we define the matrix $X=I_{d_A}\otimes I_{d_B} +
\sum x_i \sigma_i$, the dual form of our semi-definite program may be
written
 \begin{eqnarray*}
\label{sdp2}
{\rm maximize} \quad &  1-{\rm Tr}\,X \rho   \\
{\rm subject\ to} \quad & {\sf Sym'}(X\otimes {\bf I}) \geq 0.
\end{eqnarray*}
The advantage of this semi-definite program from the point of view of
analytical work is that the number of variables is very much smaller
and we will see in the following that it can be further reduced by
symmetry methods. Numerical implementations solve both optimizations
at once. To see that this program is also strictly feasible consider
the point $x_i=0$ for all $i$, for which $X=I_{d_A}\otimes I_{d_B}$
and so ${\sf Sym'}(X\otimes {\bf I})> 0$. From this we can conclude,
by Theorem 3.1 of \cite{VB:sp} as described above, that if the maximum
of this optimization is less than or equal to one (${\rm Tr}\,X_{\rm
  opt} \rho \geq 0$) there is a symmetric extension for $\rho$. It is
not necessary to construct the extension explicitly.

We now use the symmetry of the Werner states to simplify this
semi-definite program. The Werner states have the property that
$(U\otimes U) \rho_{\rm W} (U\otimes U)^{\dagger} = \rho_{\rm W}$ for
all unitary transformations $U$ on $\mathcal{H}$ \cite{EW:uuu}. As a
result the objective function is unchanged under the action
$X\rightarrow (U\otimes U)X(U\otimes U)^{\dagger}$ since ${\rm
  Tr}\,(U\otimes U) X(U\otimes U)^{\dagger}\rho_{\rm W}={\rm
  Tr}\,X(U\otimes U)^{\dagger}\rho_{\rm W} (U\otimes U)= {\rm
  Tr}\,X\rho_{\rm W}$. Similarly if we choose an $X$ such that the
positivity constraints are satisfied then
\begin{equation*}
  \label{eq:pos}
  U^{\otimes (s_a+s_b)}
  {\sf Sym'}(X\otimes {\bf I}) U^{\dagger\otimes (s_a+s_b)}  \geq 0.
\end{equation*}
On the other hand, $U^{\otimes (s_a+s_b)}$ commutes with all matrices
that permute the tensor factors and so it commutes with the
symmetrization operation. As a result
\begin{equation*}
  \label{eq:pos2}
  {\sf Sym'}((U\otimes U)X(U\otimes U)^{\dagger}\otimes {\bf I}) \geq 0.
\end{equation*}
So the set of allowed matrices $X$ is also invariant under the
operation $X\rightarrow (U\otimes U)X(U\otimes U)^{\dagger}$. In fact
since sums of positive matrices are positive convex combinations of
matrices $(U\otimes U)X(U\otimes U)^{\dagger}$ for different $U$ are
also feasible and also achieve the same value of the objective
function. As a result we may restrict our attention to matrices
$\bar{X}$ such that $\frac{1}{{\rm Vol}(U)}\int (U\otimes U) \bar{X}
(U\otimes U)^{\dagger}dU=\bar{X}$. For any feasible $X$ of lower
symmetry there is some feasible symmetrized matrix $\bar{X}$ such that
${\rm Tr}\,\bar{X}\rho_{\rm W}={\rm Tr}\,X\rho_{\rm W}$. As discussed
in \cite{EW:uuu} all the Hermitian matrices of this form may be
written $\bar{X}=x_I I_{d_A}\otimes I_{d_B}+xV$ for some real $x_I$
and $x$. For the matrices $X$ arising in our semi-definite program
$x_I=1$ so we are left with a single variable optimization. (Note that
there are general methods along these lines for reducing the
dimensionality of semi-definite programs with symmetry
\cite{GP:sss,Rains:spde}. The key property is the {\em convexity} of
the optimization.)

The objective function to be maximized for the Werner state is $1-{\rm
  Tr}\,X\rho_{\rm W}=-x\Phi$. Since $\rho_{\rm W}$ is separable for
$\Phi\geq 0$ we already know that an extension will exist for positive
$\Phi$ and can assume that $\Phi$ is negative in the following. As a
result we wish to find the maximum value of $x$ for which the
following matrix inequality holds
\begin{equation*}
  I_d^{\otimes (s_a+s_b)} + x{\sf Sym'}(V\otimes
  {\bf I}) \geq 0.
\end{equation*}
The eigenvalues of ${\sf Sym'}(X\otimes {\bf I})$ are simply
$1+x\lambda_i$ where $\lambda_i$ are the eigenvalues of ${\sf
  Sym}(V\otimes {\bf I})$. Supposing that the largest magnitude
negative eigenvalue of ${\sf Sym'}(V\otimes {\bf I})$ is
$-\lambda_{\rm m}$, the optimum is $x_{\rm opt}=1/\lambda_{\rm m}$.
Since the optimization is now over a single variable, the
semi-definite program essentially reduces to an eigenvalue problem.
The optimum of the semi-definite program will be $-x_{\rm opt} \Phi$
and if this is less than or equal to one there is a symmetric
extension for $\rho_{\rm W}$. So $\rho_{\rm W}$ has a symmetric
extension over the range $-1/x_{\rm opt}=-\lambda_{\rm m}\leq \Phi
\leq 1$. Note that the symmetrization is a completely positive map and
maps positive matrices $X$ to positive matrices ${\sf Sym'}(X\otimes
{\bf I})$.  Since $X=I+xV$ is positive for $-1 \leq x\leq 1$, ${\sf
  Sym'}(X\otimes {\bf I})$ must also be positive over this range so
$x_{\rm opt}\geq 1$.

To proceed further we must evaluate the smallest eigenvalue of ${\sf
  Sym'}(V\otimes {\bf I})$. The matrix $V$ swaps the Hilbert spaces
belonging to Alice and Bob and therefore is an element of the
representation of the group of permutations $\mathcal{S}_{s_a+s_b}$
that is made up of permutations of the $s_a+s_b$ copies of
$\mathcal{H}$. If $\pi_{(i,j)}$ is the matrix that swaps the $i$-th
and $j$-th spaces, we have $V=\pi_{(1,2)}$.  For clarity we will
modify the order of the Hilbert spaces that we used above and imagine
that odd-numbered spaces $i=1,3,...,s_a+s_b-1$ are copies of Alice's
system and that even-numbered spaces are copies of Bob's.

As a preliminary, consider evaluating ${\sf Sym}(\sigma \otimes
I_d^{\otimes (s-1)})$ on $\mathcal{H}^{\otimes s}$. Clearly $\sigma
\otimes I_d^{\otimes (s-1)}$ is unaffected by any permutation that
fixes the first object and so in some sense the only permutations that
matter are the ones that swap the first Hilbert space with one of the
others. More formally, we can divide the elements of the permutation
group of $s$ objects into cosets of the subgroup that fixes the first
object. Any permutation can be realized by some permutation that fixes
the first object followed by the transposition $\pi_{(1,i)}$ for some
$i$ that labels which coset the permutation is in. Thus if we write
the elements of the subgroup that fixes the first element as
$\bar{\pi}$ we have
\begin{eqnarray*}
  \label{eq:symop1}
  {\sf Sym} (H)&=&\frac{1}{s!}
  \sum_{\pi}
  \pi H \pi^{\dagger} \nonumber \\
&=& \frac{1}{s} \sum_{i=1}^{s} \pi_{(1,i)} \left (
  \frac{1}{(s-1)!}\sum_{\bar{\pi}} \bar{\pi}
H \bar{\pi}^{\dagger}\right)\pi_{(1,i)}.
\end{eqnarray*}
Finally, $\sigma\otimes I_d^{\otimes (s-1)}$ maps to itself under all
$\bar{\pi}$ so ${\sf Sym}(\sigma \otimes I_d^{\otimes (s-1)})=\sum
\pi_{(1,i)} (\sigma \otimes I) \pi_{(1,i)}/s$. Applying this
observation to the local symmetrization in the particular case where
$X=V=\pi_{(1,2)}$ gives
\begin{equation*}
  \label{eq:symsimple}
  {\sf Sym'}(V\otimes {\bf I})=\frac{1}{s_a s_b}
  \sum_{i=1}^{s_a} \sum_{j=1}^{s_b} \pi_{(2i-1,2j)}.
\end{equation*}
So the symmetrized matrix reduces to a linear combination of
transpositions of Hilbert spaces.

The matrices $\pi$ form a reducible unitary representation of the
symmetric group $\mathcal{S}_{s_a+s_b}$. There is a single unitary
transformation that will block-diagonalize all the $\pi$ with the
blocks being, possibly repeated, irreducible representations of the
group (see for example \cite{book:fultonharris}). Since this is a
single unitary transformation applied to all $\pi$ it does not change
the eigenvalues of linear combinations of different $\pi$. Such linear
combinations will be positive if and only if all the blocks are
positive and so we can restrict our attention to the eigenvalues of
the blocks of ${\sf Sym'}(V\otimes {\bf I})$ in this basis. Now
suppose that the alternating representation occurs in this
decomposition at least once. Since in this irreducible representation
even permutations are represented by $1$ and odd ones by $-1$, the
value of this block of ${\sf Sym'}(V\otimes {\bf I})$ is $-1$. This
implies that $x_{\rm opt}\leq 1$. Combined with the earlier lower
bound on the optimum we have $x_{\rm opt}=1$ and $\rho_W$ has a
symmetric extension for all allowed values $-1\leq \Phi \leq 1$.

It is a standard result that the alternating representation occurs in
the decomposition into irreducibles of the representation of
$\mathcal{S}_{s_a+s_b}$ by permutations of tensor factors of
$\mathcal{H}_d^{\otimes (s_a+s_b)}$ so long as $s_a+s_b \leq d$
\cite{book:fultonharris}. In fact it is easy to see that the
alternating representation occurs if and only if there is a
non-trivial completely antisymmetric subspace of
$\mathcal{H}_d^{\otimes (s_a+s_b)}$. This completes the argument. {\it
  All Werner states have a $(s_a,s_b)$-symmetric extension and an LHV
  theory for all Bell experiments with $(s_a,s_b)$ settings if
  $s_a+s_b\leq d$.}

Note that the converse is not true, a symmetric extension may exist
even if this condition is not met. Consider $s_a=s_b=2$ and $d=2$.
In both these cases the decomposition includes the two-dimensional
irreducible representation of $\mathcal{S}_4$ \cite{book:fultonharris}
that can be generated by
\begin{equation*}
  \label{eq:irrep2}
  \tilde{\pi}_{(1,2)}=\left(
    \begin{array}{rr}
1 & 0 \\ -1 & -1
    \end{array}
\right)=\tilde{\pi}_{(3,4)},\quad \tilde{\pi}_{(2,3)}=\left(
 \begin{array}{cc}
0 & 1 \\ 1 & 0
    \end{array}
\right).
\end{equation*}
Any block corresponding to this irrep is then equal to
\begin{equation*}
  \label{eq:block2}
  \left(
  \tilde{\pi}_{(1,2)} + \tilde{\pi}_{(1,4)} + \tilde{\pi}_{(2,3)} +
  \tilde{\pi}_{(3,4)} \right)/4=\left(
 \begin{array}{rr}
0 & -\frac{1}{2} \\ -\frac{1}{2} & 0
    \end{array}
\right).
\end{equation*}
Since this has eigenvalues $(-1/2,1/2)$ we have $x_{\rm opt}\leq 2$.
It is straightforward to check the blocks corresponding to the other
irreducible representations to confirm that $x_{\rm opt}=2$.  As a
result we may say that Werner states in two dimensions have
$(2,2)$, $(2,1)$ and $(1,2)$-extensions if $-1/2\leq \Phi \leq
1$. It is
straightforward to confirm numerically that the same statement applies
to decomposable $(2,2)$-quasi-extensions. The irreducible
representations of $\mathcal{S}_n$ for large $n$ are typically high
dimensional. Note, however, that further block diagonalization is
possible in principle since ${\sf Sym'}(V\otimes {\bf I})$ is
invariant under any permutation of Alice's spaces or of Bob's. This
means that each block of ${\sf Sym'}(V\otimes {\bf I})$ is in the
commutant algebra of the reducible representation of
$\mathcal{S}_{s_a} \times \mathcal{S}_{s_b}$ that arises from
restricting an irreducible representation of $\mathcal{S}_{s_a+s_b}$
to a representation of $\mathcal{S}_{s_a} \times \mathcal{S}_{s_b}$.
This commutant algebra will typically have very much smaller block
size.

\bibliographystyle{hunsrt}
\bibliography{refs}


\end{document}